\documentclass[journal]{IEEEtran}

\usepackage{cite}

\ifCLASSINFOpdf
  \usepackage[pdftex]{graphicx}
\else
   \usepackage[dvips]{graphicx}
\fi
\usepackage[cmex10]{amsmath}
\usepackage{array}
\begin{document}
%
\title{Multiple-Access Quantum Key Distribution Networks}
%
%
%

\author{Mohsen~Razavi,~\IEEEmembership{Member,~IEEE}
\thanks{This research has received funding from the Seventh Framework Programme under Grant Agreement 277110.}
\thanks{M. Razavi is with the School
of Electronic and Electrical Engineering, University of Leeds, Leeds, LS2 9JT,
UK, e-mail: m.razavi@leeds.ac.uk.}
}

\maketitle

\begin{abstract}
This paper addresses multi-user quantum key distribution networks, in which any two users can mutually exchange a secret key without trusting any other nodes. The same network also supports conventional classical communications by assigning two different wavelength bands to quantum and classical signals. Time and code division multiple access (CDMA) techniques, within a passive star network, are considered. In the case of CDMA, it turns out that the optimal performance is achieved at a unity code weight. A listen-before-send protocol is then proposed to improve secret key generation rates in this case. Finally, a hybrid setup with wavelength routers and passive optical networks, which can support a large number of users, is considered and analyzed. 
\end{abstract}


%
\IEEEpeerreviewmaketitle

\section{Introduction}
%
%
%
%
\IEEEPARstart{C}{onsider} different branches of a bank within a metropolitan area. Suppose any two of them may need to exchange a secret key---a sequence of random bits---without trusting any other parties. They require to be immune against any eavesdropping attacks, during the key exchange, and, also, at any time in the future. Except for possibly some initial setup requirements, they also require the convenience of remotely exchanging the keys without relying on traditional meet-and-exchange protocols. Whereas most known cryptographic solutions to this problem rely on computational complexity, hence threatened by future technological advancements, there is one emerging technology that can meet all above requirements. Quantum key distribution (QKD) is a future-proof protocol, whose security results from the laws of physics as modeled by quantum mechanics \cite{BB_84, ShorPreskill_00}. This paper studies the above and similar multi-user scenarios, where QKD lends itself to network environments.

QKD relies on single-photon technology requiring us to deal with challenges of generation, transmission, and detection of single photons. A quantum leap forward was recently taken by the introduction of decoy-state protocols \cite{Lo:Decoy:2005}, which relaxed the need for generating ideal single photons by replacing them with weak laser pulses. These weak laser pulses still need to go through lossy channels with possibly additional background noise. Nevertheless, throughout the past two decades, point-to-point QKD has successfully been demonstrated over fiber and free-space channels covering distances over 100~km \cite{GYS_04, Zeilinger_Decoy_07}, and at gigahertz transmission rates over 60~km of optical fiber \cite{GHz_QKD_09}. This distance is large enough to support applications in local and metropolitan areas without relying on quantum repeaters \cite{Zoller_Qrepeater_98, Razavi.DLCZ.06, Razavi.Lutkenhaus.09, Razavi.Amirloo.10}.

Several essential steps have recently been taken to make the QKD technology available to the public. The most important breakthrough is the implementation of QKD over existing fiber-optic infrastructures \cite{MagiQ_Verizon, Telcordia_1550_1550, Telcordia_1550_1310, WDM_QKD_Han_2009, Tokyo_QKD, Townsend_QI_home_2011}. This is especially important because QKD protocols are extremely sensitive to the background noise, and the crosstalk noise between classical and quantum channels substantially deteriorates QKD performance. Nevertheless, the integration between classical and quantum communications is inevitable. It is partly because all QKD protocols rely on conventional classical communications for their operation, but, more importantly, because it will be too costly to use dedicated channels for QKD applications. All setups proposed in this paper would therefore support both types of communications on a common platform. 


\begin{figure}[!t]
\centering
\includegraphics[width = \linewidth]{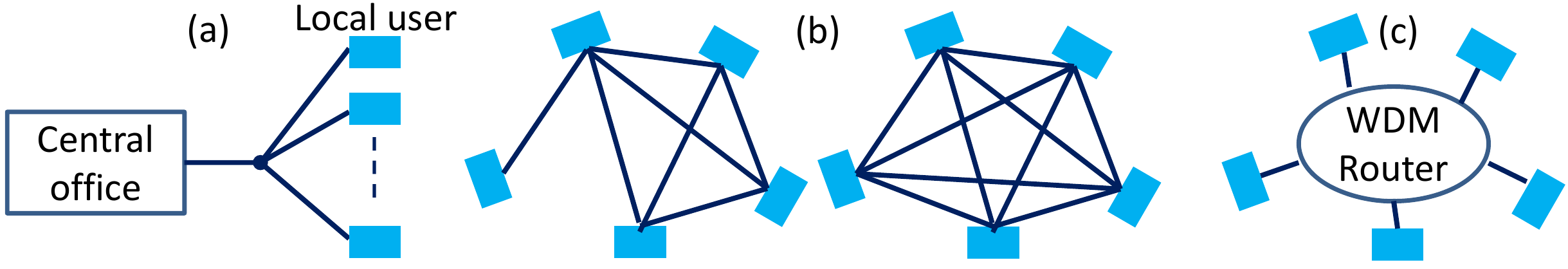}
\caption{Different configurations for QKD networks: (a) Access network, where all local users can exchange a key with the central office; (b) mesh networks, where there is a route between any pairs of users, possibly, via some other trusted users. The full mesh network, on the middle right, does not require trusting any other nodes, but its required number of links grow quadratically with the number of users; and (c) a switch-based network, where different users are connected together via a switch, such as a WDM router, and can exchange a key using designated wavelengths.}
\label{fig1}
\end{figure}

Another key requirement for a more versatile use of QKD is its expansion from point-to-point links to multi-user networks, which will be at the core of this work. The initial steps toward this end have already been taken. Key exchange between one central node and several access nodes has been studied and developed in the past few years \cite{Townsend_QI_home_2011, QKD_Network_JLT}; see Fig.~\ref{fig1}(a). A real-time demonstration of key exchange over a mesh network of several nodes was also demonstrated by the SECOQC project \cite{secoqc}, and, more recently, in Tokyo QKD network \cite{Tokyo_QKD}. Such examples, despite their being important breakthroughs in the field, suffer from a general limitation. In order to enable key exchange between any two network users, they often require to {\em trust} certain intermediate nodes. In the case of access networks, the central node must be trusted by all nodes, and, in the case of mesh networks, there is not necessarily a direct route between any two nodes, unless a full mesh network is used; see Fig.~\ref{fig1}(b). The latter architecture is too costly with the number of required channels scaling quadratically with the number of nodes. 

Instead of connecting all nodes directly together, as in a full-mesh network, one can use the concept of switching. This idea has been used in a wavelength division multiplexing (WDM) network, in which any two nodes are assigned a certain wavelength for key exchange and a wavelength router links them together \cite{WDM_QKD_Han_2009}; see Fig.~\ref{fig1}(c). Such a system, however, requires a linearly increasing number of wavelength resources with the number of nodes, even if we efficiently reuse the channels \cite{WDM_QKD_saving}. In the field demonstration of \cite{WDM_QKD_Han_2009}, there is also a specific detection module required per incoming wavelength, which makes the system even more costly.

In this paper, we look at other available dimensions, i.e, time and code, to be employed in time/code division multiple-access (TDMA/CDMA) QKD networks. They are particularly useful if combined with a WDM routing setup, in which case, each WDM node can serve as a hub through which multiple TDMA/CDMA users can be supported. In particular, a hybrid WDM-CDMA setup does not require any network-wide time coordination and is suitable if the need for key exchange is sporadic. To this end, we first study a TDMA/CDMA based multi-user star-topology QKD network, and find average lower bounds on the secret key generation rates when decoy-state protocols are used. For CDMA we use optical orthogonal codes (OOCs) to address each node \cite{OOC}. It turns out that the lower the code weight the higher the key generation rate is. It follows that a code with weight one is the best option for QKD. This case is equivalent to a TDMA system whose users are not synchronized. To get the advantage of both TDMA and CDMA, a listen-before-send (LBS) protocol is proposed, in which users attempt to pick a free time slot by first listening to the channel and then proceed to key exchange only if no other user is detected in that slot. 

The rest of the paper is organized as follows. In the next section, we review original and decoy-state QKD protocols and the existing lower bound on the secret key generation rate in the latter case. In Sec. III, we study a passive star QKD network using TDMA and CDMA as their multiple-access methods. We also calculate the average key rate for the LBS protocol. Some numerical results will be presented in Sec. IV, before introducing hybrid WDM-T/CDMA QKD networks in Sec V. Section VI concludes the paper.

\section{Quantum Key Distribution}
In this section, we first review the original BB84 protocol, that relies on ideal single photons \cite{BB_84}, and then we present its decoy-state variant and the existing lower bounds for secret key generation rate in the case of point-to-point QKD links.

The BB84 protocol enables two parties, namely, Alice and Bob, to securely exchange, or, more precisely, extend a secret key sequence. The probability of failure in achieving either an identical or insecure key, possibly due to the presence of eavesdroppers, can be made arbitrarily small. It performs this task through the following steps. Step one includes the transmission of a raw key by Alice's encoding and sending single photons to Bob. Encoding is done in two, randomly chosen, nonorthogonal polarization/phase bases, which creates maximum uncertainty in decoding if one does not know the basis used. The bases will be revealed later, via authenticated classical communications, in order that Alice and Bob turn their raw keys into sifted keys by keeping only the bits for which the same basis has been used for encoding and decoding. In the next step, Alice and Bob attempt to correct for possible discrepancies in their sifted keys using error-correction techniques. If they find the quantum bit error rate (QBER) too high, they will abort the protocol, otherwise they apply privacy amplification to their corrected keys to bring the amount of leaked information to eavesdroppers below a desired threshold. 

Since the introduction of BB84, it has been tempting to use weak laser pulses, with less-than-unity average photon numbers, instead of ideal single photons. It can, however, be shown that by taking advantage of multiple-photon components of a coherent state, an eavesdropper can obtain information about the key, hence reducing the secret key generation rate. In fact, if the channel transmissivity between Alice and Bob is denoted by $\eta$, the key generation rate will drop proportionally to $\eta^2$ if one uses coherent states, whereas it scales with $\eta$ if ideal single photons are used \cite{LutkenhausJahma_02}. It was until the advent of the decoy-state protocol that was shown that by a simple trick one could get back to the same scaling with $\eta$ even if weak laser pulses are used. The trick is in Alice, occasionally, changing her laser intensity from its typical {\em signal} value to some {\em decoy} values. These random-power pulses provide Alice and Bob with extra information that help them better detect potential eavesdroppers.

The secret key generation rate per transmitted pulse, for a BB84 protocol that uses decoy coherent states and threshold detectors for its implementation, in the limit of an infinitely long key, is lower bounded by $\max[0,P(Y_0)]$, where \cite{Lo:Decoy:2005}
\begin{equation}
\label{rate}
P(Y_0) =  1/2 \left( -f(E_\mu) Q_\mu H(E_\mu) + Q_1 [1-H(e_1)] \right),
\end{equation}
where $H(p) = -p\log_2p -(1-p)\log_2(1-p)$, for $0\leq p \leq 1$, $f(E_\mu) \geq 1$ is the error correction inefficiency,
\begin{equation}
Q_\mu = 1 - (1 - Y_0) e^{-\eta \mu} \,\,\mbox{and} \,\,
E_\mu = [e_0 Y_0 + e_d (1- e^{-\eta \mu})]/Q_\mu
\end{equation}
are, respectively, the overall gain and the QBER,
\begin{equation}
\label{Q1e1}
Q_1 = Y_1 \mu e^{- \mu} \,\, \mbox{and} \,\,
e_1 = (Y_0/2 + e_d \eta)/Y_1
\end{equation}
are, respectively, the gain and the error rate of a single-photon state, $\mu$ is the average number of photons in a signal pulse, $Y_1 = Y_0 + \eta (1-Y_0)$ is the yield of a single-photon state, $\eta$ is the total transmissivity of the link including the efficiency of Bob's detectors, $e_d$ represents the misalignment error, and $Y_0$ is the probability of a click on the Bob's side without having any incident photons from Alice. In a point-to-point link, $Y_0$, the yield of the vacuum state, models the photodetectors' dark current and the background noise.


\section{QKD over Star Networks}
\label{QKDstar}
Switching and routing devices are among key components of modern communication networks. For quantum applications, such as QKD, one should also consider whether the employed switching scheme is compatible with the single-photon regime of operation. In this section, we consider a simple, low cost, switching idea suitable for quantum optical communications. In our setup, $N$ users, at an identical distance $L$ from each other, are connected via an $N \times N$ star coupler. A star coupler combines the signals at its input ports and broadcasts them to all output ports. Each output port will then carry a shadow of each input signal. This results in a multiple-access scenario, which must be handled at the receiver. Here, we employ the two approaches of TDMA and CDMA to distinguish between different users' signals. TDMA is, in principle, interference free, but it requires network-wide time synchronization. It, however, provides us with a reference to which we can compare the performance of other proposed techniques such as CDMA. We should also bear in mind that the splitting nature of star couplers will prevent us from supporting a large number of users. This transparent architecture can, however, be used as a private local/metropolitan area network, or as the access part of a larger optical network, an example of which will be discussed in Sec.~\ref{Sec_WDM-CDMA}.

QKD relies on both classical and quantum communications for its key exchange protocol. In a multi-user scenario, we should, therefore, support both types of communications for all users. It is an ongoing research in finding out the best strategy for coexistence of weak, single-photon, quantum signals and strong, multi-million-photon, classical pulses on a single strand of fiber. The main problem is the crosstalk noise induced by classical signals over quantum channels, which can fully mask the information in the QKD single photons. Here, we use the scheme proposed in \cite{Telcordia_1550_1310} by using the 1550~nm wavelength band for classical applications and the 1310~nm band for quantum signals. Depending on the employed band demultiplexer additional filtering may also be required. In our analytical model, we model the above crosstalk effect by considering a fixed background noise in our quantum channels.

\begin{figure}[!t]
\centering
\includegraphics[width = 0.8\linewidth]{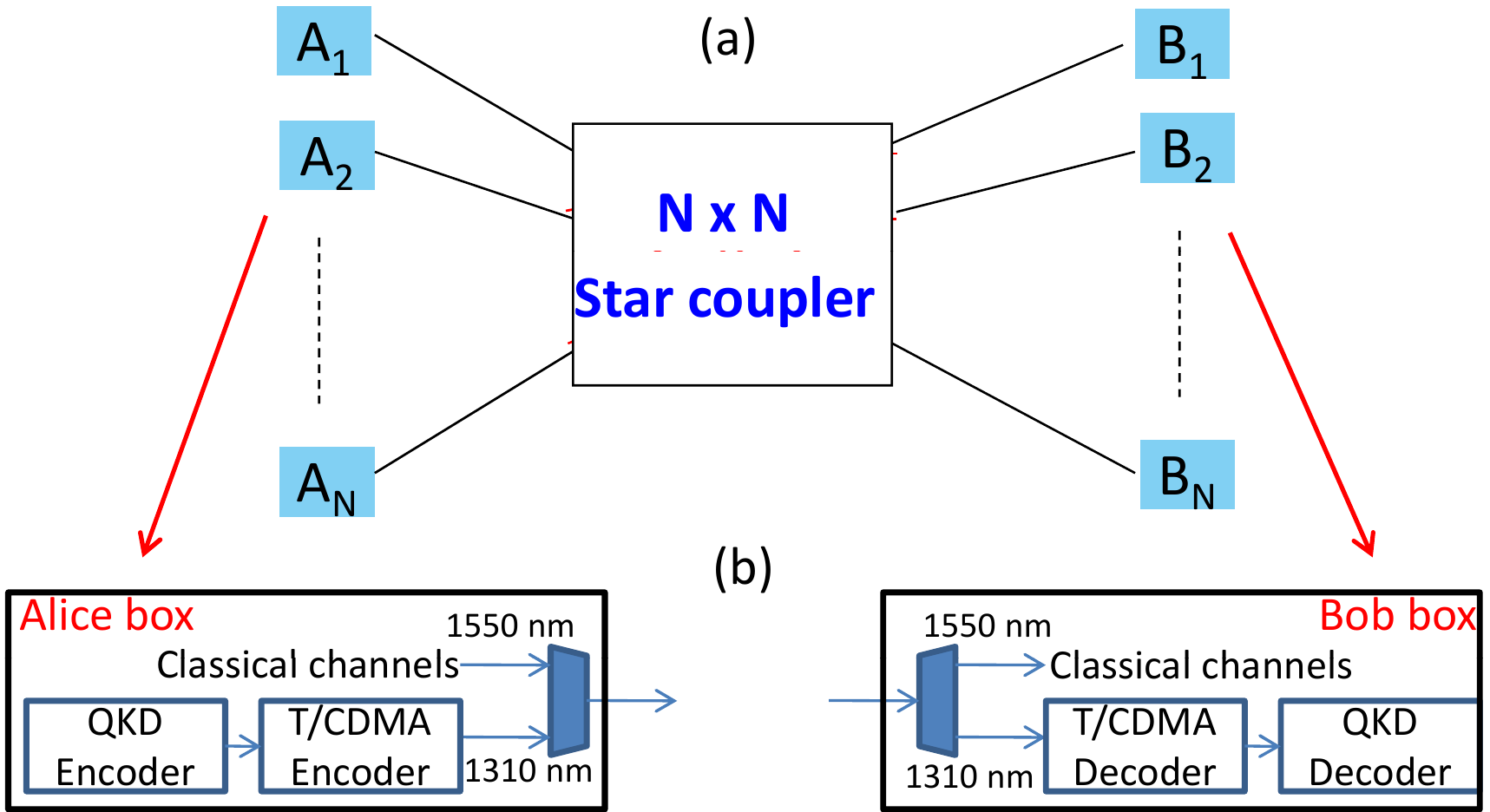}
\caption{(a) A star QKD network. All users are connected to each other via a transparent star coupler. TDMA or CDMA encoding is used to enable multiple access among quantum users. (b) Each user is equipped with an Alice box and a Bob box. Classical and quantum signals, in two different wavelength bands, are multiplexed and demultiplexed, respectively, in these boxes running independently of each other.}
\label{fig2}
\end{figure}

Each user, in Fig.~\ref{fig2}(a), has an Alice box and a Bob box; see Fig.~\ref{fig2}(b). Alice boxes include a QKD encoder using a faint laser source at the 1310~nm band, followed by a multiple-access (MA) encoder, which addresses, either in time or by a code, the intended receiver. The output of the MA encoder is multiplexed with classical channels in the 1550~nm  band. The quantum and classical parts can be run independently of each other. Bob boxes include a band demultiplexer, which separates the quantum and the classical channels, an MA decoder, which attempts to extract the intended weak laser pulse from the received signal, followed by necessary QKD measurements. Circulators can be used to enable bidirectional transmission to/from Bob/Alice boxes. 

Only one QKD measurement module, in our setup, is needed per user. This is a cost-effective approach as the most expensive elements of a QKD link are commonly the avalanche photodiods (APDs) used for single-photon detection. Only one user, at a time, can, then, exchange a key with a certain Bob box. Proper media-access layer (MAC) protocols must then be used to avoid collisions, of attempting to exchange a key with the same user, between users. This can, however, be coordinated via classical channels before two QKD users start their protocol, and is assumed throughout the paper. We also assume that the users employ the decoy-state variation of the standard BB84 protocol \cite{Lo:Decoy:2005} and that channel phase/polarization distortions are compensated at the receiver.

\subsection{TDMA QKD Networks}

\begin{figure}[!t]
\centering
\includegraphics[width = 0.8\linewidth]{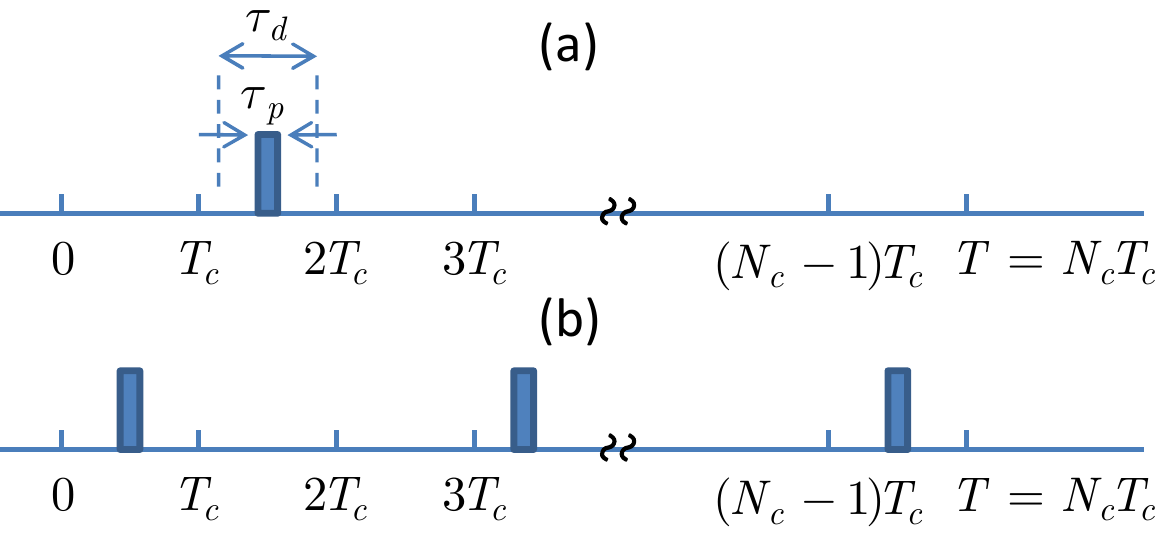}
\caption{(a) A TDMA time frame for QKD users. Each frame has a width $T$, and is divided into $N_c$ time slots (chips). The width of the laser pulse is denoted by $\tau_p$ and the gate width of single-photon detectors is denoted by $\tau_d$. (b) An example of an OOC sequence. In CDMA QKD, in order to send a raw key bit, instead of sending a single pulse, one should send a sequence of pulses corresponding to the designated code to the intended receiver.}
\label{fig3}
\end{figure}
 
In TDMA QKD networks, each receiver is assigned a certain time slot out of $N_c$ available time slots. The total frame length is denoted by $T$, which must be longer than the dead time of single-photon detectors; see Fig.~\ref{fig3}(a). The width of each laser pulse is denoted by $\tau_p$, which must not exceed $T_c \equiv T/N_c$, or the chip period. The MA encoder's task is to make sure that the transmitted weak laser pulses will arrive at the receiver at the right time. For instance, in order to send a raw key bit to user $k$, for $k=1,\ldots,N$, the transmitted signal state is as follows 
\begin{equation}
\hat \rho_k (\mu) = \sum_{n=0}^\infty{\frac{\mu^ne^{-\mu}}{n!} |n \rangle _{kk}\langle n|},
\end{equation}
where $|n \rangle_k$ represents an $n$-photon Fock state corresponding to the $k$th chip, and no photons in any other chips. The MA decoder box removes all but the signal in the desired time slot by opening the detector gate during the expected arrival time. Under these conditions, the secret key generation rate per user is lower bounded by 
\begin{equation}
\label{R_TDMA}
R_{\rm TDMA} = \max [P(Y_{\rm TDMA})/T, 0] ,
\end{equation}
where
\begin{equation}
\label{YTDMA}
Y_{\rm TDMA} = (\gamma_{\rm dc} + \eta_d \gamma_{\rm xtalk} B_{\rm opt}) \tau_d , 
\end{equation}
with $\gamma_{\rm dc}$ being the total dark count rate generated by photodetectors in a Bob box, $\eta_d$ the quantum efficiency of single-photon detectors at Bobs' boxes at 1310~nm, $\gamma_{\rm xtalk}$ the background rate of crosstalk photons leaked from classical channels at the receiver per unit of bandwidth, $B_{\rm opt}$ the optical bandwidth of the receiver, and $\tau_d \geq \tau_p$ the photodetectors' gate width. Throughout the paper, we assume that $\gamma_{\rm dc}$ and $\gamma_{\rm xtalk}$ are fixed and identical for all users. In principle, because of practical constraints on photodetectors, $\tau_d$ can become greater than $T_c$, in which case we have to consider the interference from other users as well. Here, we assume that $\tau_d \leq T_c$ so that the above TDMA rate, in (\ref{R_TDMA}), provides us with a reference in comparison with other methods. The total channel transmissivity to be used in (\ref{rate})--(\ref{Q1e1}) and (\ref{R_TDMA}) is given by
\begin{equation}
\eta = \eta_d \times 10^{-\alpha L/10} /N,
\end{equation}
where $\alpha$ is the channel loss factor in dB per unit of length. With $N_A \leq N$ {\em active} pairs of users operating in the network, the total secret key generation rate for the entire network is then given by
\begin{equation}
R_{\rm TDMA}^{\rm (tot)} = N_{A} R_{\rm TDMA} .
\end{equation}

\subsection{CDMA QKD Networks}

\begin{figure}[!t]
\centering
\includegraphics[width = \linewidth]{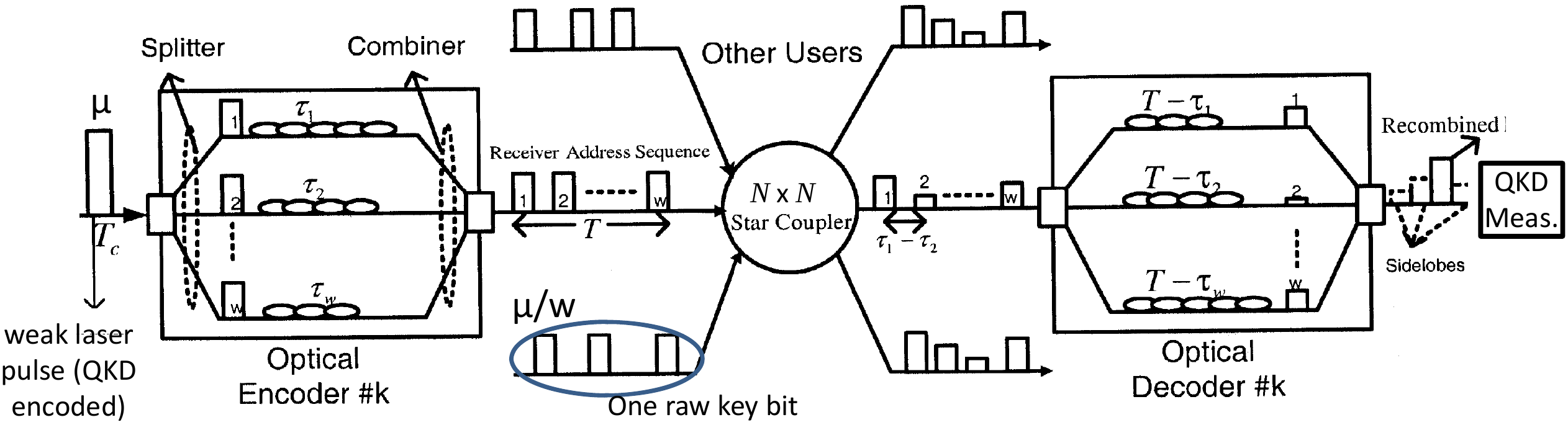}
\caption{Passive encoders and decoders for CDMA QKD. In order to reduce the total splitting loss, one can use active switches at the decoder.}
\label{fig4}
\end{figure}

In CDMA QKD networks, each receiver is assigned a certain code of length $N_c$ by which it can be addressed. In our case, we use OOCs, with minimal auto- and cross-correlation properties. Each code in an OOC family can be represented by an $N_c$-long sequence of zeros and ones. A bit one in position $k$ of a code sequence represents a pulse in the $k$th chip in Fig.~\ref{fig3}(b). Bits zero represent no pulses. In such zero-one codes, two code sequences $A$ and $B$ are overlapping whenever a pulse in code $A$ is in the same time slot as a pulse in code $B$. The minimal cross-correlation criterion guarantees that once such an overlap between two pulses occurs, there will be no other overlapping pulses in codes $A$ and $B$. That will also be the case if we shift one code sequence against the other in which case the total extent of overlapping between two codes does not exceed one pulse duration. In our QKD setup, in order to send a raw key bit to a user with an OOC sequence of weight $w$ and pulse positions $1 \leq k_1 < k_2 ,\ldots < k_w \leq N_c$, the following state is generated in each CDMA frame:
\begin{equation}
\label{CDMA_enc}
\frac{1}{w}\sum_{i=1}^w{\hat \rho_{k_i}(\mu/w)}.
\end{equation}
For such a code, the decoder box will combine the received pulses in positions $k_1$, ..., $k_w$ into one pulse and pass it to the next stage for QKD measurements; see Fig.~\ref{fig4}. In the absence of any interfering users, and assuming that different components in (\ref{CDMA_enc}) will be added incoherently \cite{Razavi_OCDMA_I}, the decoded signal pulse will then be Poisson distributed with mean $\eta \mu$.

Both the MA encoding and decoding steps in OOC CDMA can be performed transparently using passive optical elements, making them compatible to QKD applications. Figure~\ref{fig4} shows passive structures for CDMA encoders and decoders, where splitters and combiners along with proper delay elements have been used to split a single pulse into multiple pulses at the transmitter, and to recombine them again at the receiver. Note that the passive structure for the decoder results in an additional loss factor of $w$, which can be avoided if, instead of the splitter, a controllable switch is used in the decoder. The switch will direct each pulse in the incoming signal to the relevant delay branch in the decoder. For the rest of this section, we assume that such an active switching has been used in the decoder. It turns out, however, that, in the optimal regime of operation, such an assumption is not required.

A new source of background noise in a CDMA setup is the interference from other QKD users. The minimal cross-correlation property guarantees that the arbitrarily shifted versions of no two codes, belonging to an identical OOC family, will overlap at more than one pulse position. There will be, however, cases where a pulse from one code will interfere with a pulse from another code. In such a case, the background noise in the QKD channel will increase. 
Assuming $m$ such interfering users, the background noise will be given by
\begin{equation}
\label{Y_CDMA}
Y_{\rm CDMA}^{(m)} = Y_{\rm TDMA} + m \eta \mu / w ,
\end{equation}
where we assumed that all users are chip synchronous \cite{OOC}. This assumption is not required in practice, but it will simplify our analysis and help us overestimate the effect of noise in line with the lower bound in (\ref{rate}). The secret key generation rate per user, in the presence of $m$ such interfering users, is then lower bounded by 
\begin{equation}
R_{\rm CDMA}^{(m)} = \max[P(Y_{\rm CDMA}^{(m)})/T , 0] .
\end{equation}

Although, in the worst-case scenario, all active users might maximally contribute to the interference level experienced by each user, on a typical operating conditions and considering the asynchronous nature of a CDMA network, the number of interfering users follows a binomial distribution corresponding to $N_{\rm int} = N_{A} -1$ trials of a Bernoulli experiment with success probability $p=w^2/N_c$, \cite{OOC}. The collision probability $p$ represents $w^2$ cases of overlap between two codes once one shifts one code, chip-by-chip, vis. a total of $N_c$ shifts, against the other code. We can then define an {\em effective} rate of secret key generation rate per user as follows
\begin{equation}
\label{Avgrate}
\bar R_{\rm CDMA} = \sum_{m=0}^{N_{\rm int}} {B(m,N_{\rm int},p) R_{\rm CDMA}^{(m)}},
\end{equation}
where
\begin{equation}
B(m,N_{\rm int},p) \equiv \frac{N_{\rm int}!}{m! (N_{\rm int}-m)!} p^m (1-p)^{N_{\rm int}-m}
\end{equation}
is the probability of having $m$ interfering users. The total effective rate will then be given by $\bar R_{\rm CDMA}^{\rm (tot)} = N_{A} \bar R_{\rm CDMA}$.

\begin{figure}[!t]
\centering
\includegraphics[width = .7\linewidth]{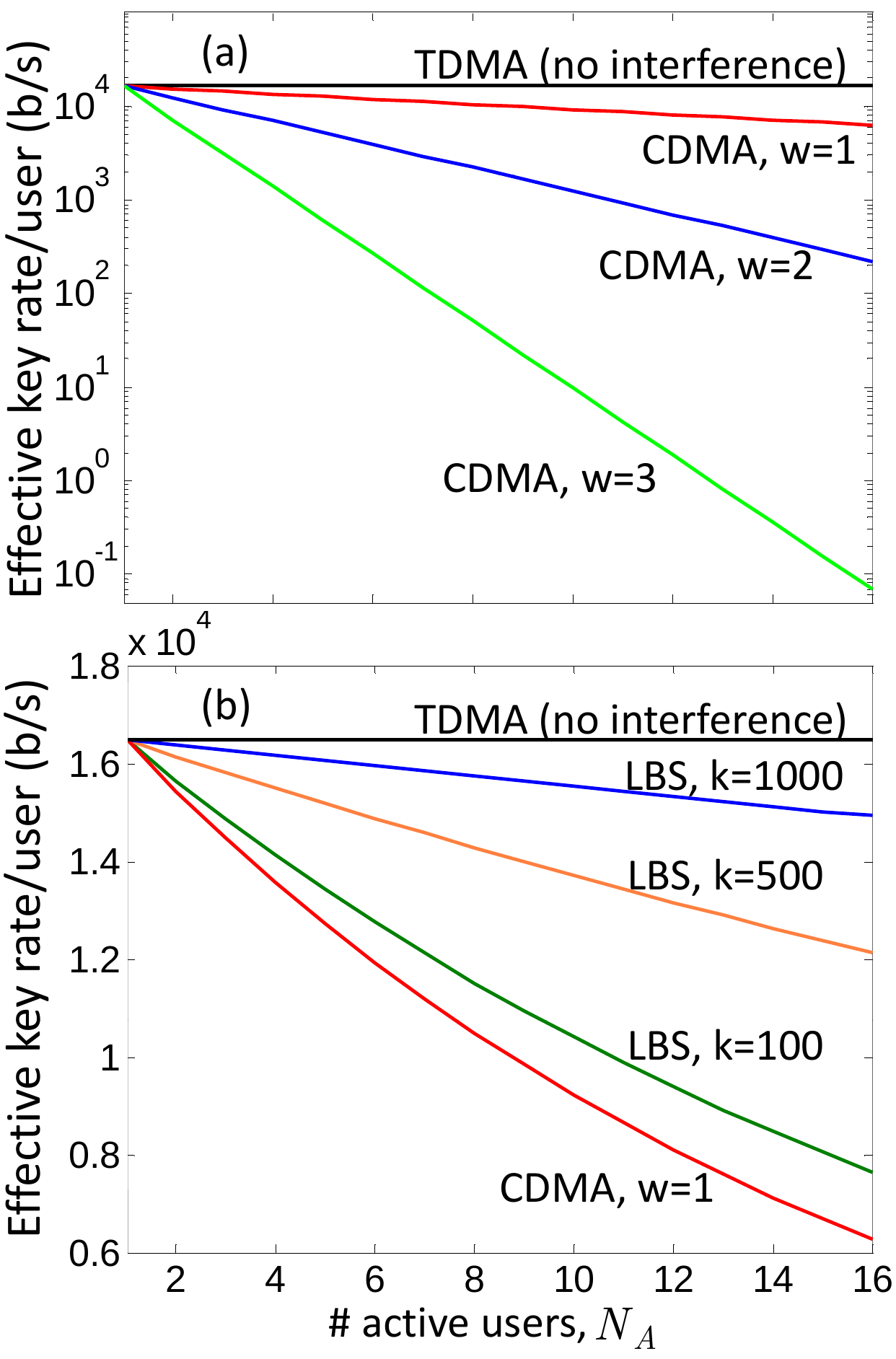}
\caption{Effective key generation rate per user versus the number of active pairs of users for (a) TDMA and CDMA QKD networks and (b) the LBS protocol. Here $w$ represents the weight of the code, and $k$ represents the number of listening periods in the LBS scheme. Nominal values used are listed in Table I.}
\label{fig5}
\end{figure}

Figure \ref{fig5}(a) shows the effective key generation rate versus the number of active users for TDMA and CDMA QKD networks. In the latter case, we have considered three different values of weight, $w$, for the code. Given that the number of available codes of length $N_c$ and weight $w$ is limited to $(N_c - 1)/w/(w-1)$, \cite{OOC}, for $N_c = 16$, a maximum of 7 and 2 users, respectively, can be supported at $w=2$ and $w=3$. In Fig.~\ref{fig5}(a), however, we have assumed that optical orthogonal codes can be assigned to all users. It is interesting to see that, whereas in classical optical CDMA an increase in the code weight would ultimately result in improving the system performance \cite{OOC2}, for QKD applications, it has the opposite effect. To see the reason for this difference, note that there are two competing terms in (\ref{Avgrate}) that affect $R_{\rm CDMA}$. The parameter $p$, for the collision probability, quadratically increases with $w$, which would require lower weight values, whereas the interference noise in (\ref{Y_CDMA}) is inversely proportional to $w$. The winner of this trade-off turns out to be the former mainly because the advantage that we obtain, by increasing $w$, from interference reduction, in a practical regime, is close to none. For low values of $w$, the interference noise, $m \eta \mu /w$, in (\ref{Y_CDMA}), is comparable with the signal rate $m \eta \mu$ at the receiver, because of which $R_{\rm CDMA}^{(m)}$ is zero for $m>0$. In fact, in order to get a nonzero value for $R_{\rm CDMA}^{(m)}$, for $m>0$, $w$ should be roughly greater than 100, which imposes impractical constraints on the system. In a practical regime of operation, where all terms in (\ref{Avgrate}), but the first, vanish, we obtain
\begin{equation}
\bar R_{\rm CDMA} = (1-w^2/N_c) ^{N_A - 1} R_{\rm TDMA},
\end{equation}
which is decreasing with $w$. The importance of collision probability in CDMA QKD networks is also noted in \cite{Razavi_QCMC_2010}.

\begin{table}[!t]
\caption{Nominal values used in the numerical results.}
\label{table_I}
\centering
\begin{tabular}{|c|c|}
\hline
Parameter & Nominal value\\
\hline
Average number of photons per signal pulse, $\mu$ & 0.48\\
Receiver dark count rate, $\gamma_{\rm dc}$ & 1E-7/ns\\
Total coupling and path loss, $10^{-\alpha L/10}$ & 6 dB \\
crosstalk rate, $\gamma_{\rm xtalk}$ & 8E-8/ns/GHz \\
Quantum efficiency, $\eta_d$ & 0.3 \\
Error correction inefficiency, $f(E_\mu)$ & 1.22 \\
Misalignment error, $e_d$ & 3.3\% \\
Star coupler size, $N$ & 16 \\
pulse/gate/chip width, $\tau_p = \tau_d = T_c = 1/B_{\rm opt}$ & 1 ns \\
Number of chips, $N_c$ & 16 \\
Frame period, $T$ & $N_c T_c$ \\
\hline
\end{tabular}
\end{table}

\subsection{LBS QKD Networks}
In the previous section, we showed that our CDMA QKD network would achieve its optimal performance at $w=1$. The case of $w=1$ corresponds to the sequences of codes with only one bit 1, representing a pulse, out of $N_c$ chips. Considering the asynchronous nature of CDMA networks, in this case, any pairs of users, who wish to exchange a secret key, can effectively choose a random chip for their QKD pulses. The performance of this scheme would be inevitably lower than that of TDMA, because, in the case of CDMA, collisions between pulses from different users are also allowed. This collision probability can, however, be reduced by employing an LBS algorithm. By using the LBS scheme, we achieve the simplicity of an asynchronous system, such as CDMA, with a performance approaching that of TDMA, as we show below. 

In our LBS QKD setup, each pair of users who need to exchange a key first pick a random chip for their pulse position. The receiver, then, listens to the channel for $k$ periods, and if no photon is detected during the chosen chip, the two users proceed with the key exchange protocol. Otherwise, they repeat the same process again, until they find a free time slot. This scheme is very similar to the well-known carrier-sense multiple access scheme, with the distinction that, in a QKD setup, we have to sense signals as weak as single photons.

The collision probability in an LBS scheme can be approximated as follows. Suppose two users $A$ and $B$ are exchanging a key by sending QKD pulses at a certain time slot. Under chip-synchronous conditions, the probability that another pair, after following the LBS protocol, choose to use the same time slot is given by $p' = [1- Y_{\rm CDMA}^{(1)}]^k/N_c$. The effective secret key generation rate can then be approximated by
\begin{eqnarray}
\label{RLBS}
\bar R_{\rm LBS} &\approx& \sum_{m=0}^{N_{\rm int}} {B(m,N_{\rm int},p') R_{\rm CDMA}^{(m)}} \nonumber \\
&\approx& (1-p')^{N_A-1} R_{\rm TDMA},
\end{eqnarray}
which approaches $R_{\rm TDMA}$ as we let $k \rightarrow \infty$.

Figure \ref{fig5}(b) shows how key generation rate improves when the number of listening periods increases. With the nominal values used in Table I, it takes on the order of 1000 listening periods to achieve the same performance as the TDMA scheme. For the network at its full capacity, that is when 16 active pairs of users are present, it takes each pair roughly 1 second to generate 15kb of secret key. The extra $1000T = 16\mu$s overhead for the LBS scheme is then negligible compared to the time needed to generate the key. Using CDMA, with $w=1$, only 6~kb/s of secret key can be generated, whereas the effective rate for TDMA is about 16.5~kb/s per user.

\section{Numerical Results}
In this section, we study the performance of the system against variations in the length of the code, the number of users, and the distance, among other parameters. We consider the three cases of TDMA, CDMA with $w=1$, and LBS with $k=500$ for our comparative study, where the effective key generation rates can, respectively, obtained from (\ref{R_TDMA}), (\ref{Avgrate}), and (\ref{RLBS}). The nominal values used are listed in Table I, which are based on practical values affordable by the today's technology. The average number of photons $\mu = 0.48$ for signal photons maximizes the secret key generation rate for CDMA ($w=1$) and TDMA QKD networks. The 6~dB path loss corresponds to a metropolitan area with roughly 20-30~km in diameter. The chosen crosstalk rate is based on the numerical values reported in \cite{Telcordia_1550_1310}. The designated dark count and quantum efficiency are also achievable by single-photon APDs at the 1300~nm band. 

\begin{figure}[!t]
\centering
\includegraphics[width = .7\linewidth]{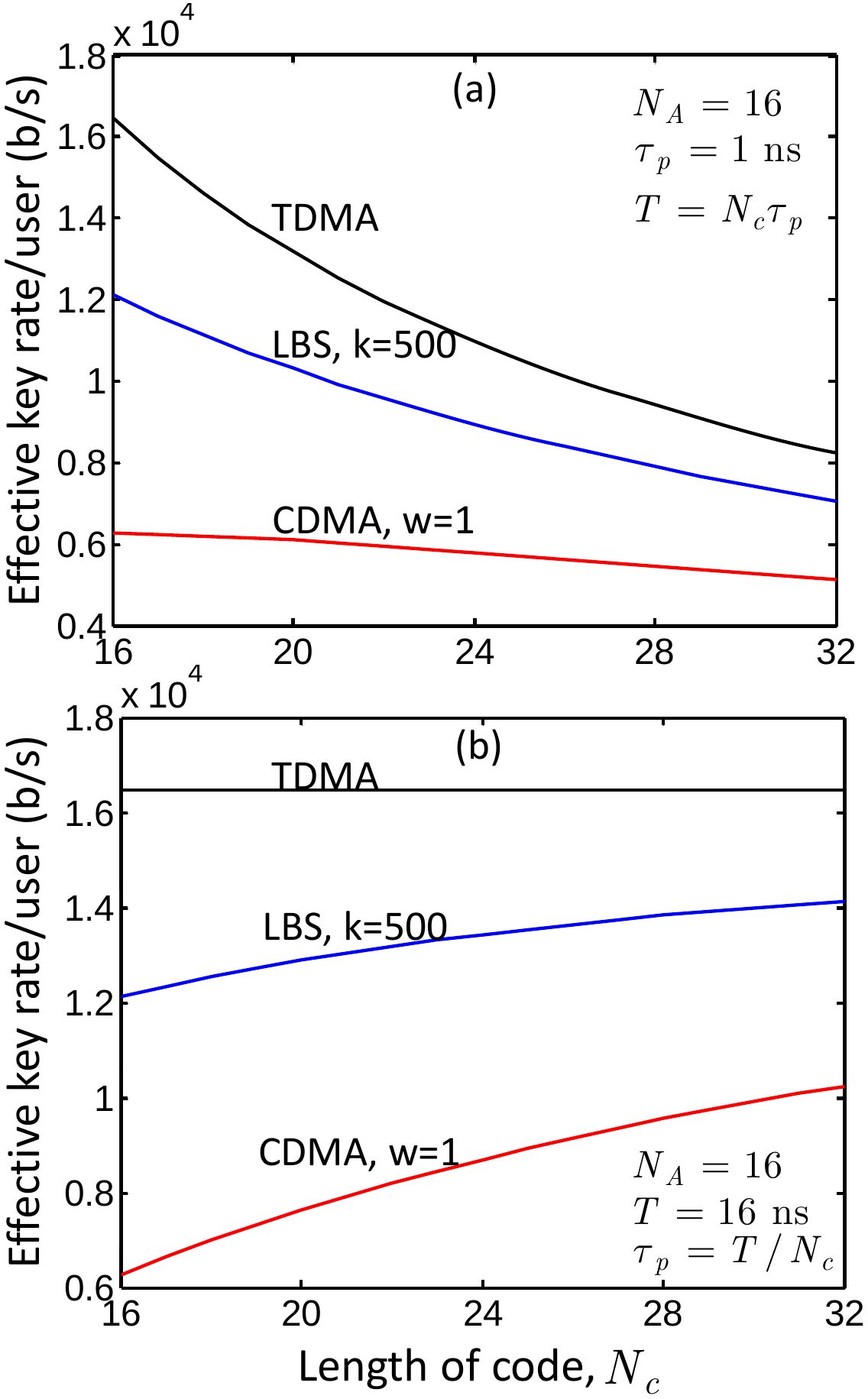}
\caption{Effective key generation rate per user versus code length for TDMA, CDMA ($w=1$), and LBS ($k=500$) QKD networks. (a) The pulse width is fixed and the period is proportional to code length. (b) The period is fixed and the pulse width is inversely proportional to code length. In both graphs, $N_A = 16$ and other parameters are taken from Table I.}
\label{fig6}
\end{figure}

Figure~\ref{fig6} depicts the effective secret key generation rate per user versus code length. Two cases have been considered. In Fig.~\ref{fig6}(a), we have assumed that the chip duration $\tau_c$ is fixed at 1~ns, and we have increased the length of the code, or, equivalently, the number of chips, $N_c$. By doing so, the total frame period $T=N_c \tau_c$ would increase and that would reduce the effective rate, inversely proportional to $T$, in the case of TDMA. By increasing $N_c$, the collision probability, in the case of CDMA, would, however, decrease, and this effect would, to some extent, balance the reduction in rate due to increase in $T$. The LBS curve lies somewhere between that of CDMA and TDMA curves. In Fig.~\ref{fig6}(b), however, the frame period $T$ is fixed at 16~ns, and increasing $N_c$ implies using shorter chips. In this case, the TDMA rate remains constant as it represents, effectively, a point-to-point system with a repetition rate of $1/T$. The CDMA and LBS performances, however, improve as a result of reduction in the collision probability. We can effectively make the chip duration as short as our detectors allow.

Based on the above results, to push the performance of the system to its maximum, for a given photodetector with time resolution $\tau_d$ and dead time $\tau_D$, one can choose $\tau_p = \tau_c = \tau_d$, and, correspondingly, to minimize the crosstalk noise, $B_{\rm opt} = 1/\tau_p$. The period $T$ will then be specified by the maximum of $\tau_D$ and $N \tau_c$. The number of chips, $N_c$, is then given by $T/\tau_c$. We use this prescription, as summarized in Table I, for the upcoming graphs.

\begin{figure}[!t]
\centering
\includegraphics[width = .7\linewidth]{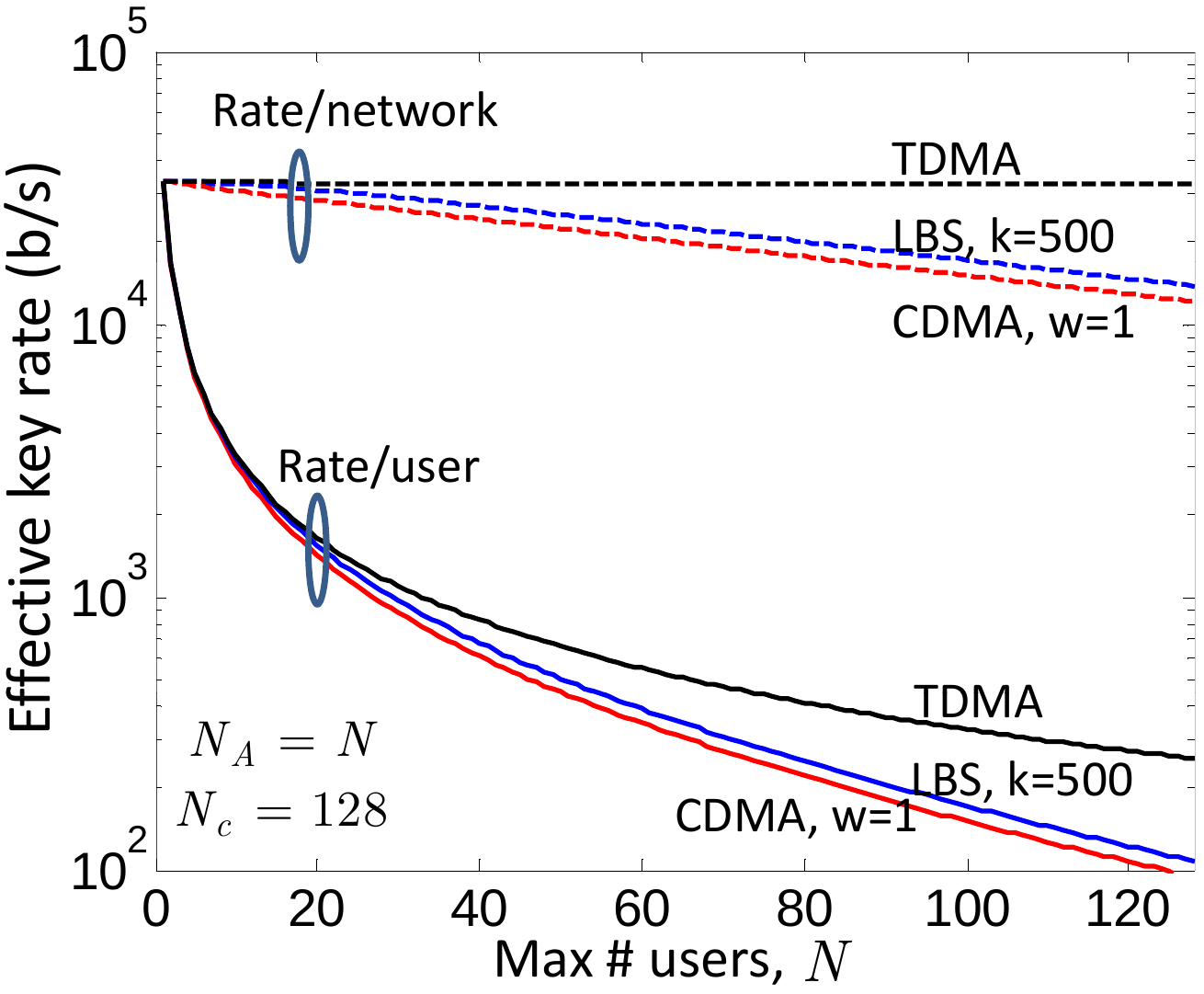}
\caption{Effective key generation rate as a function of maximum number of users, specified by the size of the star coupler switch, for TDMA, CDMA ($w=1$), and LBS ($k=500$) QKD networks. Solid lines represent rates per user, whereas dashed lines represent the total secret key generation rate in the network. It is assumed that the number of active users is at its maximum possible $N_A = N$, and the code length is fixed at 128 in all cases. Other parameters are taken from Table I.}
\label{fig7}
\end{figure}

Figure \ref{fig7} shows the effect of the splitting loss, or equivalently the maximum number of users supported by the network, on the secret key generation rate. As expected, by increasing the size of the star coupler, the splitting loss would increase and that proportionally reduce the effective rate for each pair of users. At its maximum capacity, however, when all users are paired up to exchange secret keys, the total rate remains almost constant, as shown by the dashed line in Fig.~\ref{fig7}. This implies that the total number of key bits distributed in the network is almost identical to the number of key bits generated in a point-to-point system with no splitting loss. This fair distribution of keys among users has been achieved with no loss in the case of TDMA, and small penalties in the case of CDMA or LBS schemes.

\begin{figure}[!t]
\centering
\includegraphics[width = .7\linewidth]{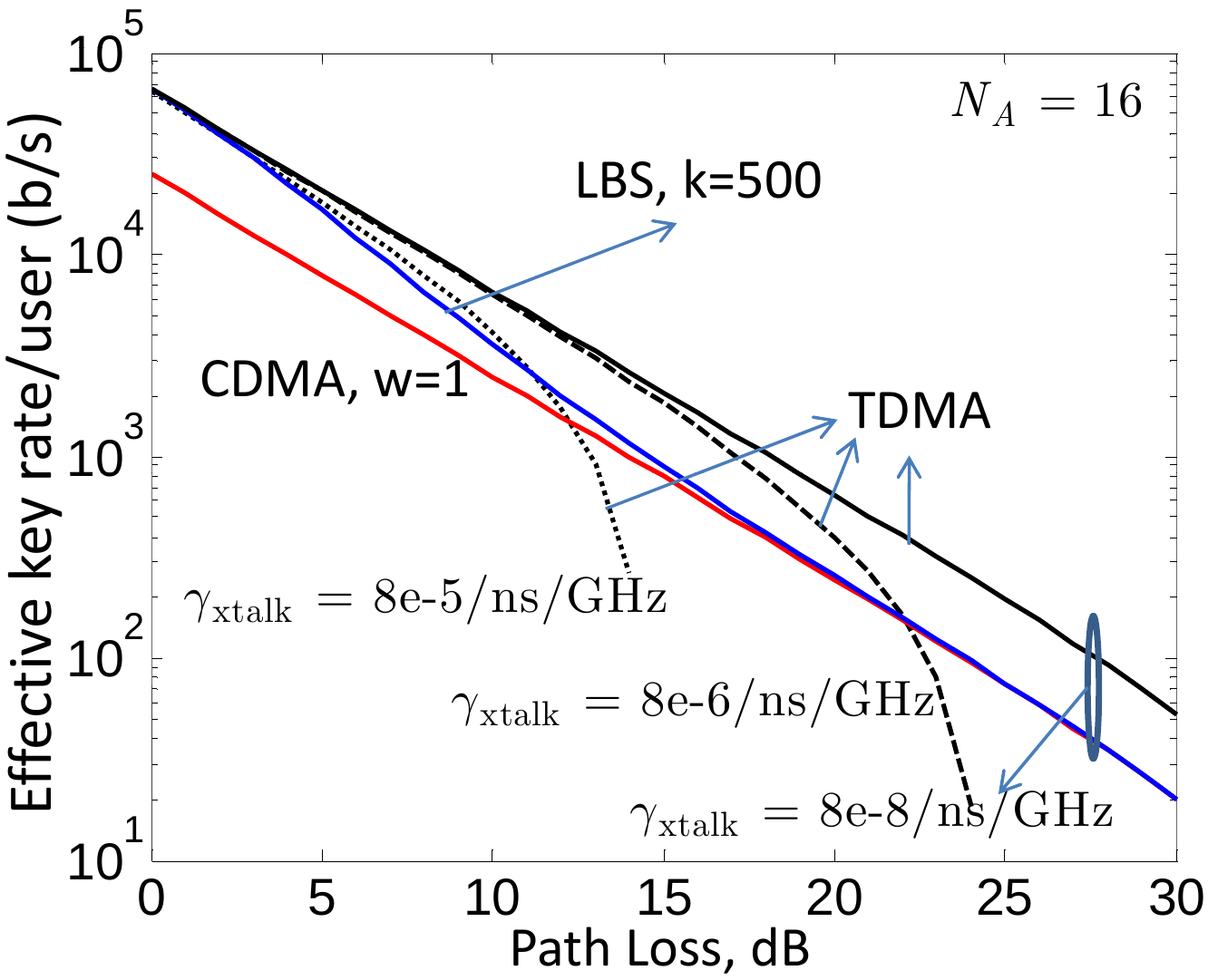}
\caption{Effective key generation rate as a function of path loss for TDMA, CDMA ($w=1$), and LBS ($k=500$) QKD networks. At $N_A = N = 16$ and $\gamma_{\rm xtalk} = 8E-8/ns/GHz$, all systems can tolerate over 30~dB of path loss (excluding the splitting effect). At higher crosstalk levels (dashed and dotted lines) the security distance drops to roughly 50~km, which is still sufficiently large for metropolitan area networks. Nominal values are taken from Table I.}
\label{fig8}
\end{figure}

Finally, Fig.~\ref{fig8} presents the effect of path loss on the effective key generation rate for the three schemes of interest. It can be seen that by increasing the path loss, the LBS protocol approaches the CDMA one, as it becomes harder and harder to detect a single photon in only $k=500$ listening periods. Another important factor shown in Fig.~\ref{fig8} is the dependence of the key generation rate on the crosstalk noise. It can be seen that even for crosstalk rates three orders of magnitude higher than the nominal value in Table I, our multiple-access system is capable of supporting 16 users at 10~dB channel loss. This makes this approach promising for home users in a passive optical network (PON) setup in future generations of optical networks. In next section, we propose a setup that combines the routing capabilities in WDM networks with the ease of access in PON systems to support a large number of QKD users.

\section{Hybrid WDM-T/CDMA QKD Networks}
\label{Sec_WDM-CDMA}

The passive star networks described in Sec.~\ref{QKDstar} support quantum and classical communications for a moderate number of users offering certain key features. They enable secret key exchange between any pairs of users without requiring them to trust any other nodes. The transparent nature of star couplers also enable coexistence of classical and quantum signals over the same infrastructure. The proposed passive structure can then be made compatible to PON architectures for consumer users. Finally, whereas the repetition rate in a point-to-point QKD link is commonly limited by the after-pulse effect in single-photon APDs, in our proposed schemes, the detectors' dead time is efficiently split between multiple users, either synchronously by using TDMA, or asynchronously by using CDMA or LBS schemes. To support a large number of users, however, we need to bring in another degree of freedom, vis. the wavelength.

\begin{figure}[!t]
\centering
\includegraphics[width = \linewidth]{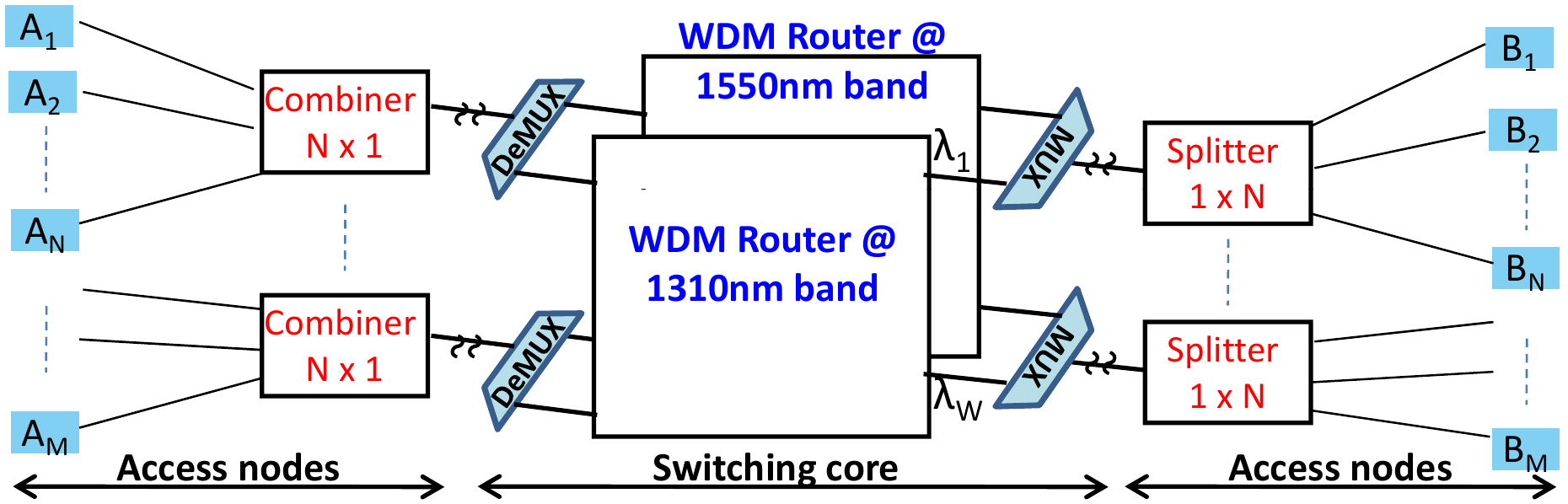}
\caption{An expansion of a hybrid WDM-T/CDMA setup. Alice and Bob boxes are the same as that of Fig.~\ref{fig2}(b). A total of $M = N \times W$ users are being supported by such a setup, where each $N$ users represent a PON with a single wavelength, and $W$ is the number of available channels. Different users of a PON are being separated in time, code, or by the LBS method. The WDM routers must provide a clear path between the intended users. MUX and DeMUX, respectively, represent band multiplexer and demultiplexer that combine/separate classical and quantum channels. In the above setup, the two systems can be run independently of each other, while the classical system can be used to coordinate, at/above the physical layer, between quantum users.}
\label{fig9}
\end{figure}

Figure \ref{fig9} illustrates a possible way for network expansion by combining WDM routing with each of TDMA/CDMA/LBS techniques. Here, Alice and Bob boxes are similar to those of Fig.~\ref{fig2}(b), except that Alice boxes now require tunable lasers. Each Bob box is designated by one of the $W$ wavelength channels in the 1310~nm band as well as a time slot corresponding to the employed TDMA/CDMA/LBS scheme. Classical communications can be independently handled, by a separate switch, in the 1550nm band. They can, nevertheless, be used to coordinate between quantum users. All Bob boxes with the same wavelength are accessed via a $1 \times N$ splitter, similar to a PON system  \cite{Townsend_PON_2007}. The WDM router combines a maximum of $N$ signals on the same wavelength and directs them to the corresponding output port. Proper MAC layer protocols will be used to avoid/reduce collisions. In essence, the network in Fig.~\ref{fig9}, for each wavelength, resembles a passive network as in Fig.~\ref{fig2}(a). The same rate analysis, with slight changes to incorporate the crosstalk noise from quantum users on different wavelengths, will then apply to this new setup as well. The switching technology at the core network determines the extent of this new form of cross talk. With quantum network at its full capacity, the background rate in (\ref{YTDMA}) will be modified to
\begin{equation}
\label{YWTDMA}
Y_{\rm TDMA} = (\gamma_{\rm dc} + \eta_d \gamma_{\rm xtalk} B_{\rm opt}) \tau_d + (W-1) \eta \mu \alpha_{\rm xt}, 
\end{equation}
where $\alpha_{\rm xt}$ represents the crosstalk factor between different wavelength channels, assumed identical for all wavelengths. Note that this assumption is in line with the lower bound nature of (\ref{rate}) as, in many cases, the only significant crosstalk terms are from adjacent channels. 

\begin{figure}[!t]
\centering
\includegraphics[width = .7\linewidth]{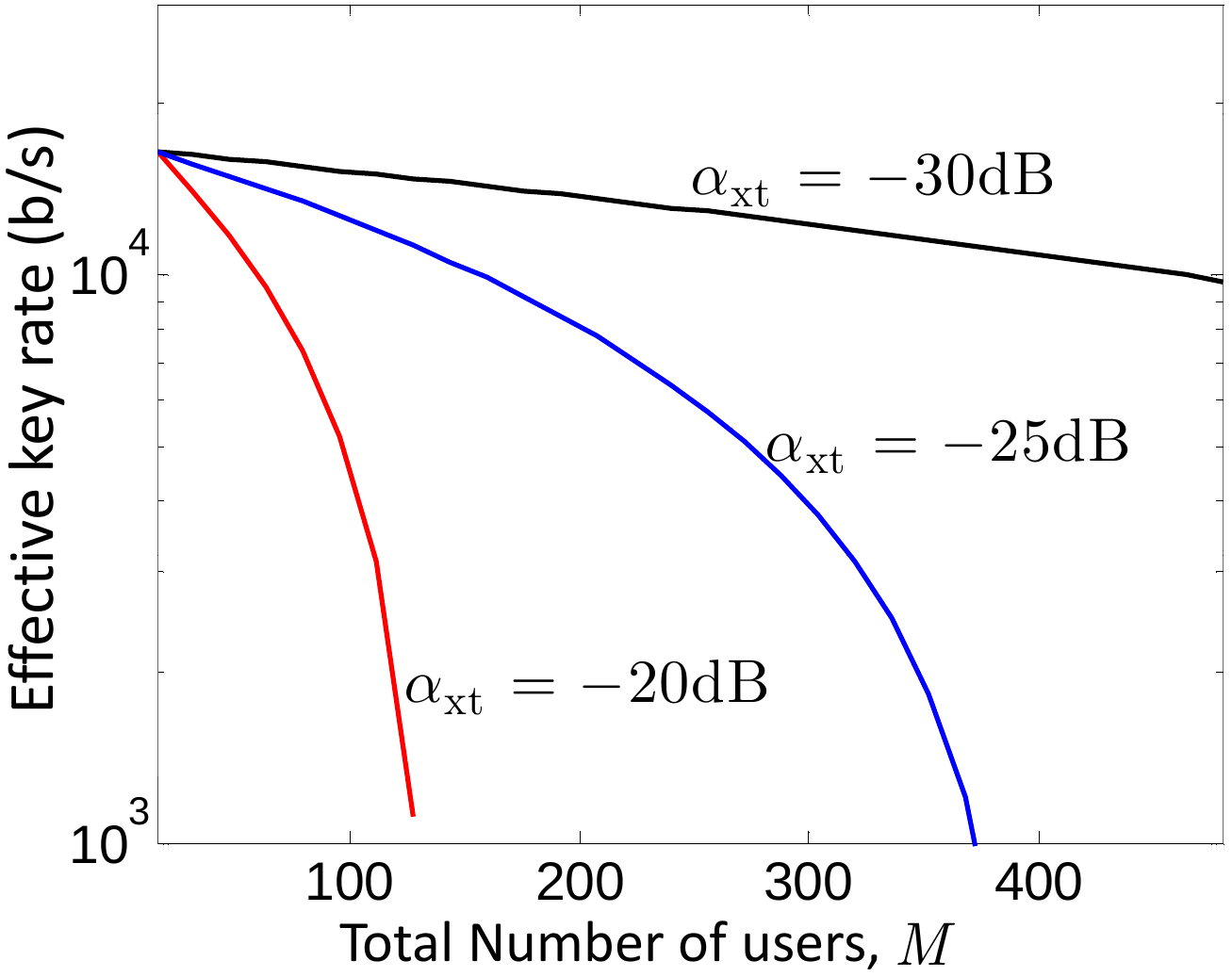}
\caption{Secret key generation rate per user for a WDM-TDMA network, as in Fig.~\ref{fig9}, versus the total number of users. Rates are calculated at the network's full capacity using nominal values in Table I for TDMA subnetworks. Different values of intra-channel crosstalk factor, $\alpha_{\rm xt}$, are considered.}
\label{fig10}
\end{figure}

Using (\ref{YWTDMA}), along with (\ref{R_TDMA}), (\ref{Avgrate}), and (\ref{RLBS}), we can obtain the secret key generation rate for hybrid WDM-TDMA/CDMA/LBS systems. Figure~\ref{fig10} shows the secret key generation rate per user, for the network in Fig.~\ref{fig9}, at its full capacity, for the WDM-TDMA scheme. We use the nominal values given in Table I for the subnetworks, and the horizontal axis represents the total number of users $M = NW$, where, here, $N = 16$. As shown in Fig.~\ref{fig10}, an isolation factor of 30~dB is sufficient to support hundreds of QKD users. With 20~dB isolation, the total number of channels that can be supported drops to 8. In principle, however, we can always use additional optical filters in Fig.~\ref{fig9} to ensure that the crosstalk noise from WDM quantum users is below the desired level.

\section{Conclusions}

In this paper, well-known techniques in classical multiple-access optical communications were applied to quantum cryptography applications. That enabled multiple users to exchange secret keys, via an optical network, without trusting any other nodes. The proposed setups offered key features that would facilitate their deployment in practice. In all of them, classical communications services were integrated with that of quantum on a shared platform, which would substantially reduce the cost for public and private users. More generally, by sharing network resources among many users, the total cost per user would shrink, making the deployment of such systems more feasible. Another cost-saving feature in our setups was their relying on only one QKD detection module per user. The setups considered were inspired by existing optical access networks as well as future all-optical networks. A passive star-coupler network was first studied when multiple QKD users could pair up and simultaneously exchange secret keys via the network. Each user could independently use classical communications as well. Different users could be distinguished in time, using a TDMA scheme, or in the code space using OOC CDMA. It turned out that, whereas TDMA QKD could offer an interference free, or, effectively, a point-to-point QKD service, CDMA QKD should deal with the interference effect. It was shown that the optimal performance for a CDMA QKD system could be achieved if codes with weight one were used, for which interference probability would be minimum. In this case, the CDMA system was similar to the TDMA one, except that no time coordination was needed between all network users. To enjoy the benefits of both TDMA and CDMA systems, a listen-before-send protocol was proposed, whose performance could approach the TDMA QKD once the number of listening periods was sufficiently large. To support a larger number of users, hybrid WDM-TDMA/CDMA architectures, potentially compatible to future all-optical networks and PON access systems, were proposed and their performance in terms of secret key generation rates and numbers of users was studied.


%




\section*{Acknowledgment}

The author would like to thank X. Ma for his fruitful discussions.

\ifCLASSOPTIONcaptionsoff
  \newpage
\fi

\end{document}